\documentclass[preprint,showpacs,preprintnumbers,amsmath,amssymb,prl]{revtex4}
\usepackage{graphicx}
\usepackage{dcolumn}
\usepackage{bm}
\usepackage[breaklinks=true,pdfborder={0 0 0}]{hyperref}

\usepackage{tabularx}

\usepackage{color}
\usepackage{tikz}

\begin{document}
\newlength{\LL} \LL 1\linewidth

\title{On the origin of the giant SHE in Cu(Bi) alloys}

\author{Dmitry V. Fedorov}
\email{dfedorov@mpi-halle.mpg.de}
\affiliation{Max Planck Institute of Microstructure Physics, Weinberg 2,
06120 Halle, Germany}
\affiliation{Institute of Physics, Martin Luther University Halle-Wittenberg,
06099 Halle, Germany}
\author{Christian Herschbach}
\affiliation{Max Planck Institute of Microstructure Physics, Weinberg 2,
06120 Halle, Germany}
\affiliation{Institute of Physics, Martin Luther University Halle-Wittenberg,
06099 Halle, Germany}
\author{Sergey Ostanin}
\affiliation{Max Planck Institute of Microstructure Physics, Weinberg 2,
06120 Halle, Germany}
\affiliation{Institute of Physics, Martin Luther University Halle-Wittenberg,
06099 Halle, Germany}
\author{Ingrid Mertig}
\affiliation{Max Planck Institute of Microstructure Physics, Weinberg 2,
06120 Halle, Germany}
\affiliation{Institute of Physics, Martin Luther University Halle-Wittenberg,
06099 Halle, Germany}

\author{Martin Gradhand}
\affiliation{H.~H.~Wills Physics Laboratory,
University of Bristol, Bristol BS8 1TL, United Kingdom}

\author{Kristina Chadova}
\affiliation{Department of Chemistry, Physical Chemistry,
Ludwig-Maximilians University Munich, Germany}
\author{Diemo K\"odderitzsch}
\affiliation{Department of Chemistry, Physical Chemistry,
Ludwig-Maximilians University Munich, Germany}
\author{Hubert Ebert}
\affiliation{Department of Chemistry, Physical Chemistry,
Ludwig-Maximilians University Munich, Germany}

\date{\today}

\begin{abstract}
Two years after the prediction of a giant spin Hall effect for the dilute Cu(Bi) alloy [Gradhand \emph{et al.},
Phys. Rev. B {\bf 81}, 245109 (2010)], a comparably strong effect was measured in thin films of Cu(Bi) alloys
by Niimi \emph{et al.} [Phys. Rev. Lett. {\bf 109}, 156602 (2012)]. Both theory and experiment consider
the skew-scattering mechanism to be responsible, however they obtain opposite sign for the spin Hall angle.
Based on a detailed analysis of existing theoretical results, we explore differences between theory and experiment.
\end{abstract}

\maketitle
One of the most interesting phenomena related to the field of spintronics is the spin Hall effect
(SHE)~\cite{Dyakonov71,Hirsch99}. It provides the opportunity to create spin currents in nonmagnetic
materials avoiding injection from a ferromagnet. For practical applications, materials with a large spin Hall angle
(SHA), the efficiency of charge to spin current conversion, are desirable. The first measurement of the giant SHE
was realized in Au with a SHA of 0.11~\cite{Seki08}. Recently, a giant SHA of $-0.12$ to $-0.15$ was measured
in highly resistive $\beta$-Ta~\cite{Liu12}, in accordance with a qualitative prediction based on a tight-binding
model for bcc Ta~\cite{Tanaka08}. Comparably large SHA's were predicted for Au(C)~\cite{Gradhand10_2}
and for Cu(Bi)~\cite{Gradhand10_3} dilute alloys from first-principles calculations. For thin films of Cu(Bi) alloys
the giant SHE was recently confirmed experimentally~\cite{Niimi12}. However, the sign of the measured spin Hall angle
($-0.24$) is opposite to the \emph{ab initio} result ($0.08$), although in both studies skew scattering
at substitutional Bi impurities is assumed to be the origin of the considered effect.

In this paper we provide an analysis of theoretical and experimental results and conclude that the sign
of the SHE measured in thin film Cu(Bi) alloys cannot be explained by the conventional skew scattering
at substitutional Bi impurities in Cu bulk. Our study is based on first-principles calculations using
the semiclassical Boltzmann equation~\cite{Gradhand10_2} and the quantum mechanical Kubo-St\v{r}eda
formula~\cite{Lowitzer11}. In addition, we present an extended version of a relativistic phase shift model
used in Ref.~\onlinecite{Niimi12}. We demonstrate that this model applied to the considered phenomenon provides
good agreement with the \emph{ab initio} calculations.

The sign of the SHA is a subtle point since different sign conventions for the spin Hall conductivity (SHC) are used
in literature. This complicates a comparison between various approaches. One definition uses the SHC in units of
the charge conductivity with the corresponding prefactor of $e^2$~\cite{Fert81,Fert11,Gradhand10_2}. Its advantage
is the coherent treatment of spin and charge conductivities providing the dimensionless spin Hall angle as their ratio.
In addition, for materials like copper with spin expectation values of the Bloch states close to one (in units of
$\hbar /2$)~\cite{Gradhand09}, the two current model can be employed. Within this model, the charge and spin Hall current
densities are given by $j_x = j_x^+ + j_x^- = \sigma_{xx} E_x = (\sigma_{xx}^+ + \sigma_{xx}^-) E_x$ and
$j_y^s = j_y^+ - j_y^- = \sigma_{yx}^s E_x = (\sigma_{yx}^+ - \sigma_{yx}^-) E_x$, respectively. Here ``$+$'' and ``$-$''
denote the two spin channels contributing to the charge conductivity $\sigma_{xx}$ and the spin Hall conductivity $\sigma_{yx}^s$
as linear response functions to an applied electric field $\bm{E} = (E_x, 0, 0)$. Although this appears natural within
the semiclassical theory~\cite{Fert81,Fert11,Gradhand10_2}, the most common definition is related to the Kubo
theory~\cite{Sinova04,Guo05,Yao05}. Here, the SHC has the prefactor of $(-e)(\hbar/2)$ replacing the electron charge
$(-e)$ by the spin units $\hbar/2$. Clearly, such a definition provides opposite sign in comparison to the first one.
Finally, one can use the SHC expressed in units of the charge conductivity but keeping the sign from the common definition
of the Kubo formula~\cite{Sinova04,Guo05,Yao05}. This was done in Ref.~\onlinecite{Lowitzer11} exploiting
the Kubo-St\v{r}eda formula. Throughout this paper the SHC, denoted as $\sigma_{\rm H}^s$,
will refer to $\sigma_{yx}^s$ of Refs.~\onlinecite{Gradhand10_2} and \onlinecite{Gradhand10_3}, $\sigma_{xy}^s$
of Ref.~\onlinecite{Lowitzer11}, and $\frac{2e}{\hbar}\sigma_{xy}^s$ of Refs.~\onlinecite{Sinova04,Guo05,Yao05}.
Taking into account that for the systems with both time and space inversion symmetry the relation $\sigma_{xy}^s = - \sigma_{yx}^s$
is valid, this procedure provides a consistent treatment of different approaches.
Obviously, the sign of the SHC determines the sign of the spin Hall angle $\alpha = \sigma_{\rm H}^s / \sigma_{xx}$ used
to quantify the SHE. This quantity is perfectly suited for the skew-scattering mechanism where $\alpha$
is independent of the impurity concentration~\cite{Gradhand10_2,Lowitzer11}.

After these introductory comments, let us compare experimental and theoretical results. The negative value of the SHA
was measured for Cu(Bi) alloys, while a positive sign of the SHE was reported for the Cu(Ir) alloy and for pure Pt
(see Fig.~2 of Ref.~\onlinecite{Niimi12}). The intrinsic contribution to $\sigma_{\rm H}^s$ calculated within
\emph{ab initio} approaches~\cite{Lowitzer11,Guo08} confirms a positive value for Pt.
In addition, it is commonly assumed that the SHE in Pt is related to the intrinsic mechanism, since reproducible
experimental results are in good agreement with the corresponding theoretical predictions. Moreover, the extrinsic
contribution was shown to be small for this material~\cite{Lowitzer11,SPIN}. Thus, the experimental values reported in
Ref.~\onlinecite{Niimi12} are given for $\sigma_{\rm H}^s$ as defined above. This point is additionally confirmed
by first-principles calculations performed for the dilute Cu(Ir) alloy. Considering the skew-scattering mechanism,
we obtain $\alpha = 0.035$ and $\alpha=0.029$ from the Boltzmann equation and the Kubo-St\v{r}eda formula, respectively,
while the experimental value is $0.023$~\cite{Niimi12}. For this alloy, both the charge and spin resistivities show almost
perfect linear dependence on the impurity concentration up to 12~at.\%~\cite{Niimi11,Niimi12}. This indicates the dominance
of the skew-scattering mechanism for the SHE in Cu(Ir). By contrast, for the Cu(Bi) alloy the experimental results deviate
from the linear dependence above 0.5~at.\% impurity concentration~\cite{Niimi12}. To handle this problem, lower concentrations
were chosen for the measurement of the reported negative SHA, assuming the skew-scattering mechanism to be dominant in this region.
This assumption was supported by numerical calculations within a resonant scattering model~\cite{Niimi12}. However, it is in
contradiction to the result of the first-principles calculations~\cite{Gradhand10_3}. To clarify this issue, we present
a derivation of an extended phase shift model.

The semiclassical approach in spherical band approximation provides the following expression for the conductivity tensor
of a crystal
\begin{equation}\label{eq.:sigma}
\begin{array}{ll}
\hat{\sigma} = \frac{e^2}V \sum\limits_\mathbf{k} \delta ({\cal E}_\mathbf{k} - {\cal E}_F)
\mathbf{v}_\mathbf{k} \circ \boldsymbol\Lambda_\mathbf{k} =
\frac{e^2 m_e k_F}{\hbar^2 (2\pi)^3} \int {\rm d} \Omega_\mathbf{k}\ 
\mathbf{v}_\mathbf{k} \circ \boldsymbol\Lambda_\mathbf{k}\ ,
\end{array}
\end{equation}
evaluated with
\begin{equation}\label{eq.:relations}
\begin{array}{ll}
{\cal E}_\mathbf{k} = \frac{\hbar^2 k^2}{2 m_e}\ ,\ \ 
\frac 1V \sum\limits_\mathbf{k} \delta ({\cal E}_\mathbf{k} - {\cal E}_F) \to 
\frac{m_e k_F}{\hbar^2 (2\pi)^3} \int {\rm d} \Omega_\mathbf{k}\ .
\end{array}
\end{equation}
Here $V$ is the system volume and $\int {\rm d} \Omega_\mathbf{k}$ refers to an integration over the angular part
of the crystal momentum $\mathbf{k}$. The mean free path in Eq.~(\ref{eq.:sigma}) is given by the Boltzmann
equation~\cite{Mertig99}
\begin{equation}\label{eq.:Lambda}
\begin{array}{ll}
\boldsymbol\Lambda_\mathbf{k} = \tau_\mathbf{k} ( \mathbf{v}_\mathbf{k} + \sum\limits_{\bf k'}
P_{\mathbf{k} \gets \mathbf{k'}} \boldsymbol\Lambda_\mathbf{k'} )\ ,
\end{array}
\end{equation}
where the momentum relaxation time $\tau_\mathbf{k}$ is defined as
\begin{equation}\label{eq.:tau}
\begin{array}{ll}
\frac 1{\tau_\mathbf{k}} = \sum\limits_{\bf k'} P_{\mathbf{k'} \gets \mathbf{k}} =
\frac{2\pi}{\hbar} c_i N \sum\limits_{\bf k'}
|T_{\mathbf{k'} \gets \mathbf{k}}|^2 \delta ({\cal E}_\mathbf{k} - {\cal E}_\mathbf{k'})
\end{array}
\end{equation}
and $\mathbf{v}_\mathbf{k} = \hbar \mathbf{k} / m_e$ is the group velocity. The microscopic transition probability
$P_{\mathbf{k'} \gets \mathbf{k}}$ describes the rate of scattering from an initial state $\mathbf{k}$ into a final
state $\mathbf{k'}$. This quantity is defined by the corresponding transition matrix $T_{\mathbf{k'} \gets \mathbf{k}}$
and scales with the impurity concentration $c_i$ and the total number of atoms $N$ in the system~\cite{Mertig99}.
This scaling holds for the dilute limit of noninteracting scatterers valid for impurity concentrations less than a few at.\%.

The derivation presented below is based on a relativistic scattering theory within the spherical band approximation,
as considered in Ref.~\onlinecite{Strange}. Following this approach, the transition matrix for the spin-conserving and
spin-flip scattering can be obtained as~\cite{Supplementary}
\begin{equation}\label{eq.:Tkk_FR}
\begin{array}{ll}
T_{{\bf k'} \gets {\bf k}}^{+ \gets +} = - \frac{8\pi^2\hbar^2}{m_e k_F V} \sum\limits_{lm}
\left(Y_l^m ({\bf \hat{k}})\right)^*~Y_l^m ({\bf \hat{k'}}) \\ \times \left[ \left(\frac{l+m+1}{2l+1}\right)
e^{i \delta_{l + \frac 12}} \sin{\delta_{l + \frac 12}} + \left(\frac{l-m}{2l+1}\right)
e^{i \delta_{l - \frac 12}} \sin{\delta_{l - \frac 12}} \right]
\end{array}
\end{equation}
and
\begin{equation}\label{eq.:Tkk_FR_SF}
\begin{array}{ll}
T_{{\bf k'} \gets {\bf k}}^{- \gets +} = - \frac{8\pi^2\hbar^2}{m_e k_F V} \sum\limits_{lm}
\left(Y_l^m ({\bf \hat{k}})\right)^*~Y_l^{m+1} ({\bf \hat{k'}}) \\ \times 
\frac{\sqrt{(l-m)(l+m+1)}}{2l+1} \left[ e^{i \delta_{l + \frac 12}}
\sin{\delta_{l + \frac 12}} - e^{i \delta_{l - \frac 12}} \sin{\delta_{l - \frac 12}} \right]\ ,
\end{array}
\end{equation}
respectively. Here $\delta_{l \pm 1/2}$ are the phase shifts for the relativistic quantum number
$j=l \pm 1/2$~\cite{Strange}. Similar to Ref.~\onlinecite{Fert11}, further on we will use the isotropic
relaxation time approximation $\tau_\mathbf{k} \approx \tau_0 = const$. Then, the relaxation time can be
easily obtained assuming $\tau_0 = \tau_{\mathbf{k}_0}$, where $\mathbf{k}_0 = (0, 0, k_F)$.
In this case we obtain~\cite{Supplementary}
\begin{equation}\label{eq.:tau_FR}
\begin{array}{rr}
\frac 1{\tau_0} = \frac{4\pi\hbar c_i}{m_e k_F V_0} \sum\limits_l \left[ (l+1) \sin^2 {\delta_{l+\frac 12}} +
l \sin^2 {\delta_{l-\frac 12}} \right] \ ,
\end{array}
\end{equation}
according to Eqs.~(\ref{eq.:relations}), (\ref{eq.:tau}), (\ref{eq.:Tkk_FR}), and (\ref{eq.:Tkk_FR_SF}).
Here $V_0 = V / N$ is the unit cell volume.

With respect to the Hall conductivity caused by the skew-scattering mechanism, the first term in Eq.~(\ref{eq.:Lambda})
is unimportant since only the \emph{scattering-in} term (vertex corrections) contributes to this
quantity~\cite{Gradhand10_2,Lowitzer11}. Moreover, only the antisymmetric part
$P_{\mathbf{k} \gets \mathbf{k'}}^{\rm antisym} = (P_{\mathbf{k} \gets \mathbf{k'}} - P_{\mathbf{k'} \gets \mathbf{k}}) / 2$
of the microscopic transition probability provides a nonvanishing contribution~\cite{Fert81,Fert11}.
In addition, we will use the approximation $\boldsymbol\Lambda_\mathbf{k'} \to \tau_\mathbf{k'} \mathbf{v}_\mathbf{k'}$
for the scattering-in term in Eq.~(\ref{eq.:Lambda}), as was done in Ref.~\onlinecite{Fert11}.

The presence of both time and space inversion symmetry provides the following relations between the two spin channels:
$\sigma_{xx}^+ = \sigma_{xx}^-$ and $\sigma_{yx}^+ = -\sigma_{yx}^-$. Thus, within the two current model discussed above,
the spin Hall angle can be written as
\begin{equation}\label{eq.:alpha_SBA}
\begin{array}{ll}
\alpha = (\sigma_{yx}^+ - \sigma_{yx}^-)/(\sigma_{xx}^+ + \sigma_{xx}^-) = \sigma_{yx}^+ / \sigma_{xx}^+ \ .
\end{array}
\end{equation}
Neglecting spin-flip transitions, the Hall component of the conductivity tensor ${\hat \sigma}^+$ is given by
\begin{equation}\label{eq.:sigma_skew}
\begin{array}{ll}
\sigma_{yx}^+ = \frac{c_i N V e^2 k_F^2 \tau_0^2}{\hbar^3 (2\pi)^5}
\int {\rm d} \Omega_\mathbf{k} \int {\rm d} \Omega_\mathbf{k'}\ k_y k_x^\prime
|T_{\mathbf{k} \gets \mathbf{k'}}^{+ \gets +}|_{\rm antisym}^2
\end{array}
\end{equation}
with $|T_{\mathbf{k} \gets \mathbf{k'}}^{+ \gets +}|_{\rm antisym}^2 = (|T_{\mathbf{k} \gets \mathbf{k'}}^{+ \gets +}|^2 -
|T_{\mathbf{k'} \gets \mathbf{k}}^{+ \gets +}|^2)/2$. Substituting Eq.~(\ref{eq.:Tkk_FR}) into Eq.~(\ref{eq.:sigma_skew}),
we obtain
\begin{equation}\label{eq.:sigma_skew_FR}
\begin{array}{ll}
\sigma_{yx}^+ = \frac{4 e^2 \hbar k_F^2 c_i}{\pi m_e^2 V_0} \tau_0^2
\left\{ \frac 19 (f_{10}-f_{12}) + \frac 15 (f_{21}-f_{23}) + \frac 27 f_{32} \right\}\ ,
\end{array}
\end{equation}
where $f_{10}$, ...~, $f_{32}$ are defined by
\begin{equation}\label{eq.:f_ll}
\begin{array}{rr}
f_{l l'} = \left\{ (l'+1) \sin{(\delta_{l'+ \frac 12} - \delta_{l + \frac 12})} \sin{\delta_{l+ \frac 12}}
\sin{\delta_{l' + \frac 12}} \right. \\ \left. + l' \sin{(\delta_{l' -  \frac 12} - \delta_{l +  \frac 12})}
\sin{\delta_{l +  \frac 12}} \sin{\delta_{l' -  \frac 12}} \right. \\ \left. - (l'+1) \sin{(\delta_{l' +  \frac 12}
- \delta_{l -  \frac 12})} \sin{\delta_{l -  \frac 12}} \sin{\delta_{l' +  \frac 12}} \right. \\ \left. -
l' \sin{(\delta_{l' -  \frac 12} - \delta_{l -  \frac 12})} \sin{\delta_{l -  \frac 12}} \sin{\delta_{l' -  \frac 12}}
\right\}/(2l'+1)\ .
\end{array}
\end{equation}
In Eq.~(\ref{eq.:sigma_skew_FR}) the contributions of $s$, $p$, $d$ and $f$ electrons are considered, neglecting terms
with $l>3$. A detailed derivation of this equation is provided in the Supplemental Material~\cite{Supplementary}.

The longitudinal conductivity in the relaxation time approximation is given by~\cite{Supplementary}
\begin{equation}\label{eq.:SigmaXX}
\begin{array}{rr}
\sigma_{xx}^+ = \frac{e^2 k_F^3}{6 \pi^2 m_e} \tau_0\ ,
\end{array}
\end{equation}
as obtained from Eqs.~(\ref{eq.:sigma}) and (\ref{eq.:relations}) neglecting the scattering-in term in the mean free
path given by Eq.~(\ref{eq.:Lambda}).

For comparison with Eq.~(2) of Ref.~\onlinecite{Niimi12}, we skip in Eqs.~(\ref{eq.:tau_FR}) and (\ref{eq.:sigma_skew_FR})
all terms with $l > 1$, assuming they are negligible for the scattering at Bi atoms. Then, using
Eqs.~(\ref{eq.:tau_FR})--(\ref{eq.:SigmaXX}), we obtain for the SHA
\begin{equation}\label{eq.:SHA}
\begin{array}{ll}
\alpha = \frac{2 \sin{\delta_0} [\sin{\delta_{1/2}} \sin{(\delta_{1/2}-\delta_0)} - 
\sin{\delta_{3/2}} \sin{(\delta_{3/2}-\delta_0)}]}{3 (\sin^2{\delta_0} + \sin^2{\delta_{1/2}} +
2 \sin^2{\delta_{3/2}})}\ ,
\end{array}
\end{equation}
where $\delta_0$ is the phase shift related to $s$ electrons ($l=0$), while $\delta_{1/2}$
and $\delta_{3/2}$ are the spin-orbit split phase shifts of $p$ electrons ($l=1$).
Equation~(\ref{eq.:SHA}) is equivalent to Eq.~(2) of Ref.~\onlinecite{Niimi12} but with opposite sign.

\begin{table}[t]
\caption{The skew-scattering contribution to the spin Hall angle $\alpha$ for the dilute Cu(Bi) alloy calculated
by means of the semiclassical and quantum mechanical \emph{ab initio} approaches as well as within the spherical band
approximation. In addition, the experimental value is given for comparison.}
\begin{tabular}{c|c}
\hline
{\bf Theory} &\ \ \ \ \ \ SHA $\alpha$\ \ \ \\
\hline
Phase shift model, Eq.~(\ref{eq.:SHA})                &   0.096  \\
Boltzmann equation~\cite{Gradhand10_3}                &   0.081  \\
Kubo-St\v{r}eda formula~\cite{Lowitzer11}             &   0.127  \\
\hline
{\bf Experiment}~\cite{Niimi12}                       &   -0.24  \\
\hline
\end{tabular}
\label{tab.:SHC}
\end{table}

The origin of this discrepancy arises from the scattering-in term of the Boltzmann equation.
In our case it is used according to Kohn and Luttinger~\cite{Kohn57}.
By contrast, Eq.~(2) of Ref.~\onlinecite{Niimi12} was based on an erroneous scattering-in
term used for the Boltzmann equation in Ref.~\onlinecite{Fert11}, that caused opposite sign
in the SHA~\cite{PrivateCommunication}.

In Table~\ref{tab.:SHC} we present the results for the skew-scattering contribution to the SHA obtained from
first-principles calculations. They are shown in comparison to Eq.~(\ref{eq.:SHA}) based on the spherical band
approximation. Clearly, the latter one provides good agreement with the Boltzmann equation. We would like to stress
that including contributions of $d$ and $f$ electrons in Eqs.~(\ref{eq.:sigma_skew_FR}) and (\ref{eq.:tau_FR})
results in almost the same value of 0.095. Thus, the assumption of Ref.~\onlinecite{Niimi12}, that the dominant
scattering process is related to $p$ electrons, is confirmed. This is in agreement with Ref.~\onlinecite{Gradhand10_3},
where it was highlighted that the spin-orbit driven scattering at Bi impurities is particularly high for $p$ electrons.
In addition, Table~\ref{tab.:SHC} demonstrates a reasonable agreement between the results obtained by the Boltzmann
equation and the Kubo-St\v{r}eda formula. According to Table~\ref{tab.:SHC}, the experimentally obtained sign of
the SHA for thin film Cu(Bi) alloys cannot be associated with the conventional skew scattering at noninteracting
substitutional Bi impurities.

Other candidates to explain the origin of the observed SHE are the intrinsic and side-jump mechanisms.
For that reason we performed corresponding calculations for Cu(Bi) alloys with different impurity concentrations
using the Kubo-St\v{r}eda formula~\cite{Lowitzer11}. Figure~1 shows the results for the SHA including the intrinsic,
side-jump and skew-scattering contribution. The sign of this quantity remains positive for the whole range of
the impurity concentrations analysed in the experiment~\cite{Niimi12}. Altogether this demonstrates
that the spin-orbit driven scattering at substitutional Bi impurities randomly distributed in bulk Cu cannot
explain the sign of the measured SHA.

\begin{figure}[t]
\label{fig.:SigmaYX_total}
\includegraphics[width=0.95\LL]{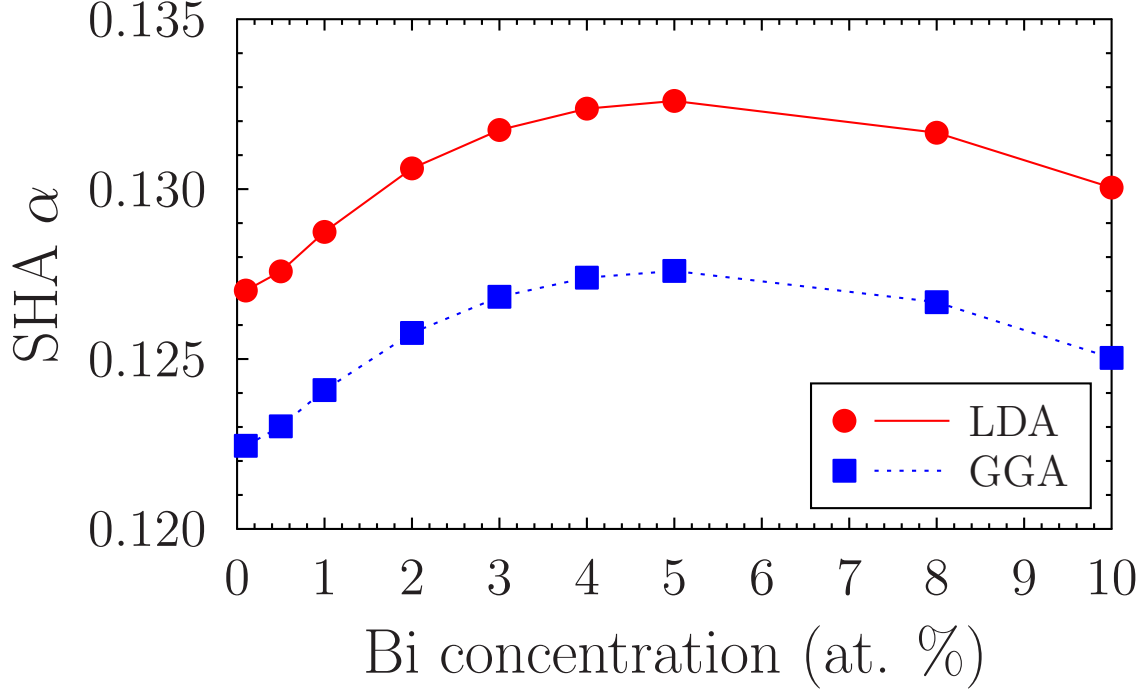}
\caption{(Color online) The spin Hall angle for the Cu(Bi) alloy with different impurity concentrations
obtained from the Kubo-St\v{r}eda formula.}
\end{figure}

A major difference between the experimental setup and the considered theoretical approaches is
that the latter ones rely on bulk materials. By contrast, the experiment is performed for thin films.
Recent \emph{ab initio} calculations have shown that Pt adatoms on fcc(111) noble metal films cause
opposite sign for the SHA than impurities within the films~\cite{Herschbach12}. We performed
corresponding calculations for ultrathin copper films with Bi impurities. Figure 2 shows the SHA
as a function of the Bi impurity position for different Cu(111) film thicknesses measured
in monolayers (ML). Here, 1ML and 2ML films provide a significantly enhanced SHA in comparison to
thicker films showing values comparable to the bulk system. As discussed in Ref.~\onlinecite{Herschbach12},
such an enhancement is caused by quantum confinement and the lack of interband transitions for the reduced
thickness. According to Fig.~2, all considered impurity positions provide positive sign for the SHA,
as in case of the bulk system. In addition, the experiment of Ref.~\onlinecite{Niimi12}
has been performed on thin film Cu(Bi) alloys with the thickness of 20~nm ($\sim$~100ML). The results
of Fig.~2 and Ref.~\onlinecite{Herschbach12} show that the influence of adatoms on the SHE seems to be
negligible for such thick films and the corresponding SHA should be close to its bulk value. However,
interface effects may still play a role due to a different geometry used in the experimental setup.
In the theory the electron spins are assumed to be along the film growth direction, while the spin current
is perpendicular to it. This is governed by the form of the conductivity tensor of Eq.~(\ref{eq.:sigma}),
which restricts the theoretical approach to in-plane charge and spin currents~\cite{Herschbach12}.
By contrast, in the experiment the spin current is injected into the Cu(Bi) film from the base Cu wire
and the spin is considered to be in plane (see Fig.~1 of Ref.~\onlinecite{Niimi12}). Finally, interface
roughness and the existence of grain boundaries in the films can provide extra scattering processes
which are not covered within the presented calculations.

\begin{figure}[t]
\label{fig.:SigmaYX_total}
\includegraphics[angle=-90,width=0.9\LL]{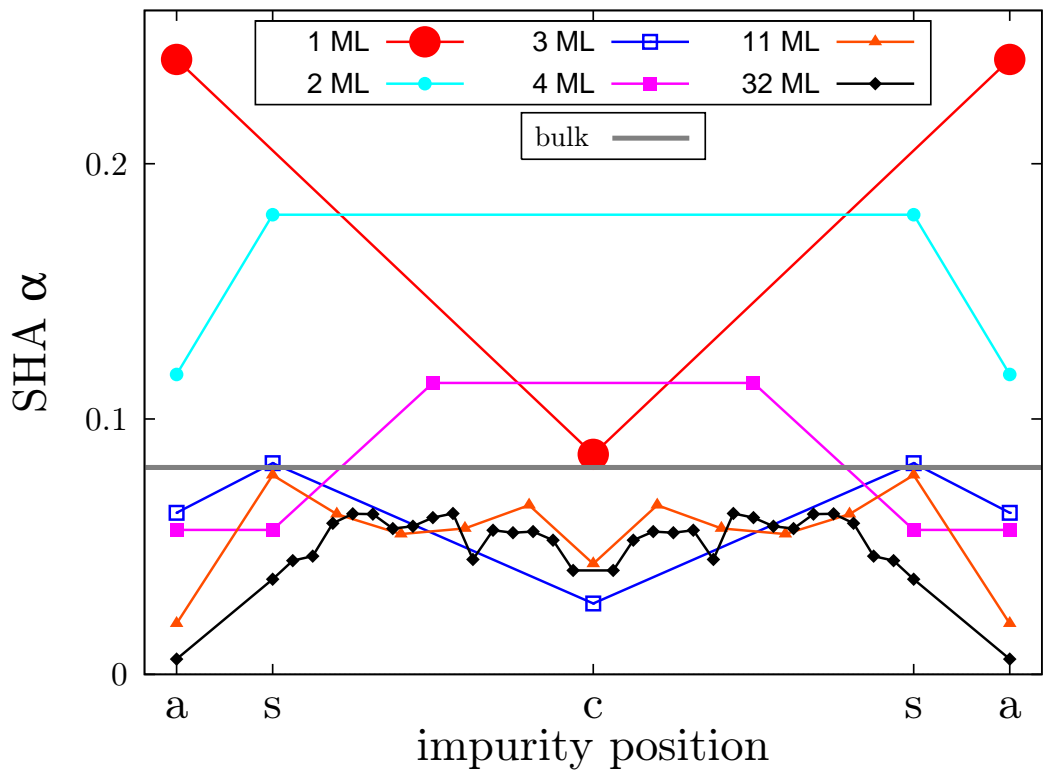}
\caption{(Color online) The spin Hall angle for ultrathin Cu(111) films with Bi impurities
which are considered at several sites shown on a normalized abscissa fixing adatom ``a'',
surface layer ``s'', and central layer ``c'' positions. The lines are to guide the eyes.}
\end{figure}

Another route to address the discrepancy between theory and experiment is related to the impurity cluster
formation which is not considered in the theory yet. Experimentally it was shown~\cite{Niimi12} that at impurity
concentration above 0.5~at.\%  Bi atoms start to segregate at the boundaries. For that reason the analysis
to extract the skew scattering contribution was restricted to lower concentrations. In this regime it was assumed
that Bi impurities are randomly distributed without short range ordering. This implies a linear relation
between the impurity concentration and the resistivity of the studied films, which was observed for lower concentrations.
However, the formation of extremely small clusters such as dimers or trimers down to lowest impurity
concentrations could not be excluded and its impact on the SHE is up to date not explored. For a description of that
case the existing theoretical approaches need to be extended. Further experimental analysis of the actual impurity
distribution is also desirable.

In addition, it has to be mentioned that within the presented theoretical
studies we have not considered effects of lattice relaxation around
Bi impurities. Although their influence seems to be relatively small for
the charge conductivity (see, for instance, comparison
between theory and experiment given by Table I in Ref.~\onlinecite{Tauber12})
they might more seriously affect the spin Hall conductivity.

In summary, we performed a detailed analysis of the giant SHE in dilute
Cu(Bi) alloys. It is based on first-principles calculations using the semiclassical
Boltzmann equation and the quantum mechanical Kubo-St\v{r}eda formula.
To elucidate the scattering contributions in terms of angular momenta,
we derived an extended phase shift model. All results of the \emph{ab initio}
and model calculations are in good agreement with respect to both sign
and magnitude of the SHA. However, the comparison with the experiment for
thin film Cu(Bi) alloys confirms the giant effect but shows disagreement
with respect to sign. Our analysis reveals that the discrepancy cannot be explained
by any calculation of the conventional skew-scattering, the corresponding side-jump
or the intrinsic mechanism. Routes to clarify the intriguing sign
problem are sketched.

\begin{acknowledgments}
We are indebted to P.~Levy, Y.~Otani and A.~Fert for valuable discussions.
The work was partially supported by the Deutsche Forschungsgemeinschaft (DFG) via SFB~762.
The authors K.C., D.K. and H.E. acknowledge support from the DFG within SFB~689 and SPP~1538.
In addition, M.G. acknowledges financial support from the DFG via a research fellowship
(GR3838/1-1).
\end{acknowledgments}

\end{document}